\documentstyle[12pt,epsfig,amssymb]{article}
\begin{document}
\begin{flushright}
Revised on \today\vspace{5mm}\\
\end{flushright}
\vspace{.5in}
\begin{center}
{\Huge MKN Theory of Bound States}\\
\vspace{.25in}
{\Large Alfred Tang}\\
\vspace{.15in}
{\em Physics Department, University of Wisconsin - Milwaukee,}\\
{\em 1900 E. Kenwood Blvd., Milwaukee, WI 53211.}
\end{center}
\vspace{5mm} \noindent

\begin{center}
{\bf Abstract}
\end{center}
\noindent
This paper derives
the MKN (Maung-Kahana-Norbury) theory of bound states which incorporates
the Lande subtraction 
technique to remove the singularities of the Cornell potential.

\newpage

\section{NRSE in Momentum Space}

Non-relativistic Schr\"{o}dinger equation (NRSE) in configuration space has 
been solved exactly for some potentials, such as the Coulomb and simple 
harmonic oscilator potentials.  NRSE with a linear potential can be solved
analytically for the $S$-state only as we will show later.  For $l>0$, we
resort to numerical methods.  NRSE $r$-space codes are commonly known
to be conditionally unstable~\cite{iitaka,succi}, while the momentum space 
codes do not have the same problem.  The momentum space code has an additional
advantage of being easily adaptable to relativistic equations.  NRSE in
momentum space takes the form
\begin{equation}
{p^{2}\over 2\mu}\,\phi({\rm\bf p}) +
\int V({\rm\bf q})\phi({\rm\bf p'})\,dp' = E\,\phi({\rm\bf p}),
\label{nrse_ms}
\end{equation}
where ${\rm\bf q}={\rm\bf p}-{\rm\bf p'}$.

\noindent Proof:

NRSE in momentum space can be derived from its configuration space
counterpart 
\begin{equation}
-{\hbar^{2}\over 2\mu}\triangledown^{2}\psi({\rm\bf x}) +
V({\rm\bf x})\psi({\rm\bf x}) = E\,\psi({\rm\bf x})
\label{nrse_cs}
\end{equation}
by Fourier transform.  First we define the following:
\begin{eqnarray}
\phi({\rm\bf p})&=&\int \psi({\rm\bf x})e^{i{\rm\bf k}\cdot{\rm\bf x}}d^{3}x,
\label{ft3}\\
\psi({\rm\bf x})&=&\int \phi({\rm\bf p})e^{-i{\rm\bf k}\cdot{\rm\bf x}}
d^{3}p,\label{ft3b}\\
{\rm\bf p}&=&\hbar{\rm\bf k},
\end{eqnarray}
and ignore factors of $2\pi$ in Fourier and inverse Fourier
transforms.  As usual, we assume
periodic boundary condition or $\triangledown\phi=\phi=0$ at infinity.
We Fourier-transform Eq.~[\ref{nrse_cs}] term by term.  The term on the
right hand side of Eq.~[\ref{nrse_cs}]
is obtained simply by Eq.~[\ref{ft3}].  The first term 
involves $\triangledown^{2}\psi$ and is transformed as
\begin{eqnarray}
&& \int \triangledown^{2}\psi({\rm\bf x})
\,e^{i{\rm\bf k}\cdot{\rm\bf x}}\,d^{3}x 
\nonumber\\
&=& \int \triangledown\psi({\rm\bf x})\,e^{i{\rm\bf k}\cdot{\rm\bf x}}\cdot
d{\rm\bf S} - \int \triangledown\psi({\rm\bf x})\cdot\triangledown
e^{i{\rm\bf k}\cdot{\rm\bf x}}\,d^{3}x \nonumber \\
&=& -i{\rm\bf k}\cdot
\int e^{i{\rm\bf k}\cdot{\rm\bf x}}\triangledown\psi({\rm\bf x})\,d^{3}x
\nonumber\\
&=& -i{\rm\bf k}\cdot\left[ \int \psi({\rm\bf x}) 
e^{i{\rm\bf k}\cdot{\rm\bf x}}\,
d{\rm\bf S} - \int \psi({\rm\bf x})
\triangledown e^{i{\rm\bf k}\cdot{\rm\bf x}}\,d^{3}x \right]
\nonumber \\
&=& -i{\rm\bf k}\cdot\left[ -i{\rm\bf k}
\int \psi({\rm\bf x}) e^{i{\rm\bf k}\cdot{\rm\bf x}}\,d^{3}x \right] 
\nonumber \\
&=& -k^{2}\phi({\rm\bf p}).
\label{ft1}
\end{eqnarray}
The second term involves $V({\rm\bf x})\psi({\rm\bf x})$.
With Eq.~[\ref{ft3b}], the second term of Eq.~[\ref{nrse_cs}] is transformed as
\begin{eqnarray}
\int V({\rm\bf x})\psi({\rm\bf x}) e^{i{\rm\bf k}\cdot{\rm\bf x}}\,d^{3}x
&=& \int V({\rm\bf x})
\left[ \int \phi({\rm\bf p'}) e^{-i{\rm\bf k'}\cdot{\rm\bf x}}\,dp' \right]
e^{i{\rm\bf k}\cdot{\rm\bf x}}\,d^{3}x \nonumber \\
&=& \int \int V({\rm\bf x})\phi({\rm\bf p'}) 
e^{i({\rm\bf k}-{\rm\bf k'})\cdot{\rm\bf x}}\,d^{3}x\,d^{3}p'
\nonumber \\
&=& \int d^{3}p' \,\phi({\rm\bf p'}) \int d^{3}x\,V({\rm\bf x})
e^{i({\rm\bf k}-{\rm\bf k'})\cdot{\rm\bf x}}
\nonumber \\
&=& \int V({\rm\bf p}-{\rm\bf p'})\phi({\rm\bf p'})\,d^{3}p'
\label{ft2}
\end{eqnarray}
At last, Eqs.~[\ref{ft3}], [\ref{ft1}] and [\ref{ft2}] are all needed to
Fourier transform Eq.~[\ref{nrse_cs}] into Eq.~[\ref{nrse_ms}].  The proof is
complete.

In this paper, we attempt to re-derive the results previously obtained by
Maung \emph{et al.} \cite{maung93} in 1993.  The power law potential in $r$ 
space is given by
\begin{displaymath}
V^{N}(r)=\left\{
\begin{array}{l l}
0 & r < 0 \\
\lambda_{N}\lim_{\eta\to 0}r^{N}e^{-\eta r} & r\geq 0, \, \eta > 0
\end{array}
\right.
\end{displaymath}

Let $G=\hbar=c=1$.  Define ${\bf q} \equiv {\bf p}-{\bf p'}$.
The momentum space potential can be obtained by Fourier transform.
\begin{eqnarray}
V^{N}({\bf q})&=&\frac{1}{(2\pi)^{3}}\int^{\infty}_{-\infty} V^{N}(r)
e^{i{\bf r}\cdot{\bf q}}d^{3}r \label{pot}\\
&=&\frac{\lambda_{N}}{(2\pi)^{3}}\lim_{\eta\to 0}\int^{\infty}_{0}\int^{1}_{-1}
\int^{2\pi}_{0}
r^{N}e^{-\eta r}e^{irq\cos\theta}r^{2}drd\cos\theta d\phi\\
&=&\frac{\lambda_{N}}{4\pi^{2}}\frac{1}{iq}\lim_{\eta\to 0}
\int^{\infty}_{0}\int^{1}_{-1}
r^{N+1}e^{-\eta r}e^{irq\cos\theta}drd(irq\cos\theta)\\
&=&\frac{\lambda_{N}}{4\pi^{2}}\frac{1}{iq}\lim_{\eta\to 0}
\int^{\infty}_{0}
r^{N+1}e^{-\eta r}\left( e^{irq}-e^{-irq} \right)dr\\
&=&\frac{\lambda_{N}}{4\pi^{2}}\frac{1}{iq}\lim_{\eta\to 0}
(-1)^{N+1}\frac{\partial^{N+1}}{\partial\eta^{N+1}}\int^{\infty}_{0}
e^{-\eta r}\left( e^{irq}-e^{-irq} \right)dr\\
&=&\frac{\lambda_{N}}{4\pi^{2}}\frac{1}{iq}\lim_{\eta\to 0}
(-1)^{N+1}\frac{\partial^{N+1}}{\partial\eta^{N+1}}
\left[ \frac{1}{\eta - iq} - \frac{1}{\eta + iq} \right]\\
&=&\frac{\lambda_{N}}{4\pi^{2}}\frac{1}{iq}\lim_{\eta\to 0}
(-1)^{N+1}\frac{\partial^{N+1}}{\partial\eta^{N+1}}
\left[ \frac{2iq}{\eta^{2} + q^{2}} \right]
\end{eqnarray}
The final form of the momentum space potential is
\begin{equation}
V^{N}({\bf q})=\frac{\lambda_{N}}{2\pi^{2}}\lim_{\eta\to 0}
(-1)^{N+1}\frac{\partial^{N+1}}{\partial\eta^{N+1}}
\left[ \frac{1}{\eta^{2} + q^{2}} \right],
\end{equation}
where $N=-1$ corresponds to the Coulomb potential and $N=1$ the linear
potential.  Together they give the Cornell potential
\begin{equation}
V({\bf q})\equiv V^{C}({\bf q}) + V^{L}({\bf q})
=V^{N=-1}({\bf q}) + V^{N=1}({\bf q}).
\end{equation}

Next we want to perform a partial wave expansion of $V^{N}$.  There are 3
useful formulas, the Wigner-Eckart Theorem\cite{sak85}
\begin{equation}
<E'l'm'|T|Elm>=\delta_{l'l}\delta_{m'm}T_{l}(E),
\end{equation}
the addition of spherical harmonics
\begin{equation}
\sum_{m} Y_{lm}(\Omega)Y^{\ast}_{lm}(\Omega')
=\frac{2l+1}{4\pi}P_{l}(\cos\theta),
\end{equation}
and the orthogonality of spherical harmonics
\begin{equation}
\int d\Omega Y^{\ast}_{lm}(\Omega) Y_{l'm'}(\Omega) = \delta_{l'l}\delta_{m'm},
\end{equation}
which are used in deriving the following result.
\begin{eqnarray}
<{\bf p}|V^{N}|{\bf p'}>&=&\sum_{lm}\sum_{l'm'}
<{\bf p}|lm><lm|V^{N}|l'm'><l'm'|{\bf p'}> \\
&=&\sum_{lm}\sum_{l'm'}<p\Omega|lm><lm|V^{N}|l'm'><l'm'|p'\Omega'> \\
&=&\sum_{lm}\sum_{l'm'}<p|<\Omega|lm><lm|V^{N}|l'm'><l'm'|\Omega'>|p'> \\
&=&\sum_{lm}<\Omega|lm><p|V^{N}_{l}|p'><lm|\Omega'> \\
&=&\sum_{lm}V^{N}_{l}(p,p')Y_{lm}(\Omega)Y^{\ast}_{lm}(\Omega') \\
&=&\sum_{l}\frac{2l+1}{4\pi}V^{N}_{l}(p,p')P_{l}(\cos\theta)
\label{eq:Vpp}
\end{eqnarray}
In scattering and bound state problems, it is customary to expand the momentum 
space wavefunction $\phi({\bf p})$ in partial waves, such that
\begin{equation}
\phi({\bf p})=\sum_{nlm}c_{nlm}\phi_{nl}(p)Y_{lm}(\Omega),
\end{equation}
where $c_{nlm}$'s are coefficients of the expansion~[p.~396 of Ref~1~].

The non-relativistic Schr\"{o}dinger equation in momentum space is given as
\begin{eqnarray}
\left( \hat{E}-\frac{{\bf p}^{2}}{2\mu} \right) \phi({\bf p})
&=&\int V^{N}({\bf q}) \phi({\bf p'}) d^{3}{\bf p'} \\
&=&\int <{\bf p}|V^{N}|{\bf p'}> \phi({\bf p'}) d^{3}{\bf p'}
\end{eqnarray}

Expand NRSE in partial waves.
\begin{eqnarray}
\left( \hat{E}-\frac{{\bf p}^{2}}{2\mu} \right) \sum_{nlm} c_{nlm} 
\phi_{nl}(p) Y_{lm}(\Omega) &=&
\int p'^{2} dp' d\Omega' \sum_{nlm} V^{N}_{l}(p,p')
Y_{lm}(\Omega) Y^{\ast}_{lm}(\Omega') \nonumber \\ 
&&\quad \times\sum_{n'l'm'} c_{n'l'm'} \phi_{n'l'}(p') Y_{l'm'}(\Omega') \\
&=&\int p'^{2}dp' \sum_{nlm} V^{N}_{l}(p,p') c_{nlm} \phi_{nl}(p') 
Y_{lm}(\Omega).
\end{eqnarray}
The $nl$-th terms can be separated by inspection with the help of the identity
$\hat{E}\,\phi_{nl}(p)=E_{nl}\,\phi_{nl}(p).$  The partial wave NRSE is
\begin{equation}
\left( E_{nl} - \frac{p^{2}}{2\mu} \right) \phi_{nl}(p) =
\int p'^{2} dp' V^{N}_{l}(p,p') \phi_{nl}(p').
\end{equation}
Use the orthogonality of Legendre polynomials, 
\begin{equation}
\int^{1}_{-1} P_{l}(x) P_{l'}(x) dx = \frac{2}{2l+1} \delta_{l'l},
\end{equation}
and Eq.~(\ref{eq:Vpp}),
we can calculate the potential matrix elements as follow.
\begin{eqnarray}
&&\quad \int^{1}_{-1} <{\bf p}|V^{N}|{\bf p'}> P_{l}(\cos\theta) d\cos\theta \\
&&=\sum_{l'} \frac{2l'+1}{4\pi} V^{N}_{l'}(p,p')
\int^{1}_{-1} P_{l}(\cos\theta) P_{l'}(\cos\theta) d\cos\theta \\
&&=\sum_{l'} \frac{1}{2\pi} V^{N}_{l'}(p,p') \delta_{l'l} \\
&&=\frac{1}{2\pi} V^{N}_{l}(p,p')
\end{eqnarray}
In order words,
\begin{equation}
V^{N}_{l}(p,p')=2\pi\int^{1}_{-1} V^{N}({\bf q}) P_{l}(\cos\theta)d\cos\theta.
\label{eq:V^N_l}
\end{equation}
Define
\begin{equation}
y \equiv \frac{p^{2}+p'^{2}+\eta^{2}}{2p'p},
\label{eq:y}
\end{equation}
and use the definition of the Legendre polynomial of the second kind 
$Q_{n}(z)$,
\begin{equation}
Q_{n}(z)={1 \over 2}\int^{1}_{-1}\frac{1}{z-t}P_{n}(t)dt,
\end{equation}
we can modify Eq.~(\ref{eq:V^N_l}) as
\begin{eqnarray}
V^{N}_{l}(p,p')
&=&2\pi\int^{1}_{-1} V^{N}({\bf q}) P_{l}(\cos\theta)d\cos\theta \\
&=&\frac{\lambda_{N}}{\pi} \lim_{\eta\to 0}
\frac{\partial^{N+1}}{\partial\eta^{N+1}}
\int^{1}_{-1} \frac{1}{q^{2}+\eta^{2}} P_{l}(\cos\theta) d\cos\theta \\
&=&\frac{\lambda_{N}}{\pi} \lim_{\eta\to 0}
\frac{\partial^{N+1}}{\partial\eta^{N+1}}
\int^{1}_{-1} \frac{1}{p^{2}+p'^{2}-2p'p\cos\theta+\eta^{2}} 
P_{l}(\cos\theta) d\cos\theta \\
&=&\frac{\lambda_{N}}{\pi} \lim_{\eta\to 0}
\frac{\partial^{N+1}}{\partial\eta^{N+1}}
\int^{1}_{-1} \frac{1}{2p'p(y-\cos\theta)} P_{l}(\cos\theta) d\cos\theta \\
&=&\frac{\lambda_{N}}{\pi} \lim_{\eta\to 0}
\frac{\partial^{N+1}}{\partial\eta^{N+1}} \frac{Q_{l}(y)}{p'p}.
\end{eqnarray}
The coulomb potential corresponds to $N=-1$ and has the form
\begin{equation}
V^{C}_{l}(p,p')=\frac{\lambda_{C}}{\pi} \lim_{\eta\to 0} \frac{Q_{l}(y)}{p'p}.
\end{equation}
The linear potential corresponds to $N=1$ and has the form
\begin{eqnarray}
V^{L}_{l} &=& \frac{\lambda_{L}}{\pi} \lim_{\eta\to 0}
\frac{\partial^{2}}{\partial\eta^{2}} \frac{Q_{l}(y)}{p'p} \\
&=& \frac{\lambda_{L}}{\pi} \lim_{\eta\to 0}
\frac{\partial}{\partial\eta} \left[ \frac{\eta}{(p'p)^{2}} Q'_{l}(y) \right]\\
&=& \frac{\lambda_{L}}{\pi} \lim_{\eta\to 0}
\left[ \frac{Q'_{l}(y)}{(p'p)^{2}} + \frac{\eta^{2}}{(p'p)^{3}} Q''_{l}(y) 
\right].
\end{eqnarray}
There are 3 useful relations in terms of Legendre polynomials of the 2nd kind
~\cite{stegun}:
\begin{eqnarray}
Q_{0}(y) &=& {1 \over 2} \ln \left| \frac{y+1}{y-1} \right|,
\label{eq:Q} \\
Q_{l}(y) &=& P_{l}(y)Q_{0}(y)-w_{l-1}(y),
\label{eq:Ql} \\
w_{l-1}(y) &=& \sum^{l}_{m=1}{1 \over m} P_{l-m}(y)P_{m-1}(y).
\end{eqnarray}

The singularities of $V^{C}_{l}(p,p')$ and $V^{L}_{l}(p,p')$ come from the 
singularies of $Q_{l}(y)$ and $Q''_{l}(y)$.  From Eq.~(\ref{eq:Ql}), it is 
obvious that the singularities of $Q_{l}(y)$ and $Q''_{l}(y)$ again come
from those of $Q_{0}(y)$, $Q'_{0}(y)$ and $Q''_{0}(y)$.  In order to treat the
singularities of the momentum space Cornell potential, we need to control the
singularities of $Q_{0}(y)$, $Q'_{0}(y)$ and $Q''_{0}(y)$ first and foremost.
Substitute Eq.~(\ref{eq:y}) into Eq.~(\ref{eq:Q}), we have
\begin{equation}
Q_{0}(y)={1 \over 2}\ln\left[ \frac{(p+p')^{2}+\eta^{2}}{(p-p')^{2}+\eta^{2}}
\right].
\label{eq:Q0}
\end{equation}
Differentiating Eq.~(\ref{eq:Q0}) yields
\begin{eqnarray}
Q'_{0}(y)&=&{1 \over 2}\frac{\partial}{\partial y} \ln \left|
\frac{y+1}{y-1} \right| \\
&=& {1 \over 2}\left[ \frac{1}{y+1} - \frac{1}{y-1} \right] \\
&=& \frac{1}{1-y^{2}} \\
&=& p'p \left[ \frac{1}{(p+p')^{2} + \eta^{2}} -
\frac{1}{(p-p')^{2} + \eta^{2}} \right] \label{eq:Q'}
\end{eqnarray}
Differentiating again gives
\begin{eqnarray}
Q''_{0}(y) &=& \frac{2y}{(1-y^{2})^{2}} \\
&=& \frac{p^{2}+p'^{2}+\eta^{2}}{p'p}
\left[ p'p \left( \frac{1}{(p+p')^{2}+\eta^{2}}
-\frac{1}{(p-p')^{2}+\eta^{2}} \right) \right]^{2},
\end{eqnarray}
or
\begin{equation}
\frac{\eta^{2}}{p'p}Q''_{0}(y)=\eta^{2} \left( p^{2}+p'^{2}+\eta^{2} \right)
\left[ \frac{1}{(p+p')^{2}+\eta^{2}}
-\frac{1}{(p-p')^{2}+\eta^{2}} \right]^{2}. \label{eq:Q''}
\end{equation}

There are two useful identities which we want to prove:
\begin{equation}
\int^{\infty}_{0} {1\over p'}Q_{0}(y, \eta=0)\, dp'
= {\pi^{2}\over 2}, \label{eq:integral_1}
\end{equation}
\begin{equation}
\int^{\infty}_{0} \left[ \frac{\eta^{2}}{p'p} Q''_{0}(y) + Q'_{0}(y) \right]
dp' = 0 \label{eq:integral_2}.
\end{equation}

\noindent
Proof: \\
The integral in Eq.~[\ref{eq:integral_1}] is derived as follow:
\begin{eqnarray}
&& \int^{\infty}_{0} {1\over p'}Q_{0}(y,\eta=0)\,dp' \\
&=& {1\over 2} \int^{\infty}_{0} {1 \over p'} \ln \left(\frac{p+p'}{p-p'}
\right)^{2} dp'\\
&=& {1\over 2} \left[ \int^{a}_{0} {1 \over x} \ln \left( \frac{x+a}{x-a} 
\right)^{2} dx + \int^{\infty}_{a} {1 \over x} \ln \left( \frac{x+a}{x-a} 
\right)^{2} dx \right] \\
&=& \int^{\infty}_{a} {1 \over x} \ln \left( \frac{a+x}{a-x}\right) dx +
\int^{\infty}_{a} {1 \over x} \ln \left( \frac{x+a}{x-a} \right) dx  \\
&=&  - \int^{0}_{\infty} {1\over ae^{-u}} \ln \left( 
\frac{a+ae^{-u}}{a-ae^{-u}} \right) ae^{-u} du +
\int^{\infty}_{0} {1\over ae^{u}} \ln \left( 
\frac{ae^{u}+a}{ae^{u}-a} \right)  ae^{u} du \\
&=&  2\left[ \int^{\infty}_{0} \ln \left( \frac{1+e^{-u}}{1-e^{-u}} \right) du
\right] \\
&=&  2\left[ \int^{\infty}_{0} \ln (1+e^{-u})\,du - 
\int^{\infty}_{0} \ln (1-e^{-u})\,du \right] \\
&=& 2\left[ {\pi^{2}\over 12} + {\pi^{2}\over 6} \right] \label{eq:sum} \\
&=& {\pi^{2}\over 2}
\end{eqnarray}
The results of Eq.~[\ref{eq:sum}] come from  relations BI((256))(10) and
BI((256))(11) in Gradshteyn and Ryzhik~\cite{GR80}.

Eq.~[\ref{eq:Q'}] has 4 simple poles: $\alpha = p+i\eta$, $\alpha^{\ast}$,
$\beta = -p+i\eta$ and $\beta^{\ast}$.  $Q'_{0}(y)$ can be rewritten as
\begin{equation}
Q'_{0}(y) = p'p \left[ \frac{-1}{(p'-\alpha)(p'-\alpha^{\ast})} +
\frac{1}{(p'-\beta)(p'-\beta^{\ast})} \right]
\end{equation}
The contour integral $\oint Q'_{0}(y) dz$ over the upper complex plane has 2 
residues: $Res(\alpha)$ and $Res(\beta)$.  Use the formula
$Res(z_{0}) = \lim_{z \to z_{0}} (z-z_{0})f(z)$ to calculate these residues,
\begin{eqnarray}
Res(\alpha) &=& \lim_{p'\to\alpha} p'p \left[ \frac{-1}{p'-\alpha^{\ast}}
+ \frac{p'-\alpha}{(p'-\beta)(p'-\beta^{\ast})} \right] \\
&=& -p(p+i\eta)\left[ {1\over 2i\eta} + 0 \right] \\
&=& -\frac{p(p+i\eta)}{2i\eta},
\end{eqnarray}
\begin{eqnarray}
Res(\beta) &=& \lim_{p'\to\beta} p'p 
\left[ \frac{-(p'-\beta)}{(p'-\alpha)(p'-\alpha^{\ast})} + 
{1\over (p'-\beta^{\ast})} \right] \\
&=& p(-p+i\eta)\left[ 0 + {1\over 2i\eta} \right] \\
&=& \frac{p(-p+i\eta)}{2i\eta},
\end{eqnarray}
\begin{eqnarray}
Res(\alpha)+Res(\beta) &=& -\frac{p(p+i\eta)}{2i\eta} + 
\frac{p(-p+i\eta)}{2i\eta} \\
&=& -\frac{p^{2}}{i\eta}.
\end{eqnarray}
Since the contour at infinity is zero and $Q'_{0}(y)$ along the real axis 
is symmetric around the origin, we obtain
\begin{eqnarray}
\int^{\infty}_{0} Q'_{0}(y)\, dp' &=& {1\over 2}\oint Q'_{0}(y)\, dz \\
&=& {1\over 2}\left( 2\pi i\sum Res \right) \\
&=& -\frac{\pi p^{2}}{\eta}.
\end{eqnarray}
$(\eta^{2}/p'p)Q''_{0}(y)$ has the same poles as $Q'_{0}(y)$ but of order 2.  
Residues of order $m$ are calculated by the formula 
\begin{equation}
Res(z_{0})=\lim_{z \to z_{0}}{1\over (m-1)!}\,
\frac{d^{m-1}}{dz^{m-1}}(z-z_{0})^{m}f(z).
\end{equation}
Again we can simplify the algebra by rewriting Eq.~[\ref{eq:integral_2}] as
\begin{equation}
\frac{\eta^{2}}{p'p} Q''_{0}(y)=\eta^{2} \left( p^{2} + p'^{2} + \eta^{2} 
\right) \left[ \frac{-1}{(p'-\alpha)(p'-\alpha^{\ast})} +
\frac{1}{(p'-\beta)(p'-\beta^{\ast})} \right]^{2}.
\end{equation}
The residues are
\begin{eqnarray}
Res(\alpha) &=& \lim_{p'\to\alpha}{d\over dp'}
\eta^{2} \left( p^{2} + p'^{2} + \eta^{2} \right) 
\left[ \frac{-1}{(p'-\alpha^{\ast})} +
\frac{p'-\alpha}{(p'-\beta)(p'-\beta^{\ast})} \right]^{2} \\
&=& \eta^{2}\left\{ 2(p+i\eta)\left[ \frac{-1}{2i\eta} \right]^{2} + 
4p(p+i\eta)\left[ \frac{-1}{2i\eta} \right]
\left[ \frac{1}{(2i\eta)^{2}} + 
{1\over 2p(2p+2i\eta)} \right]\right\} \\
&=& \frac{\eta^{2}}{2}\left\{ -{p\over\eta^{2}} - {i\over\eta} +
{p^{2}\over i\eta^{3}} + {p\over\eta^{2}} - {1\over i\eta} \right\} \\
&=& {p^{2}\over 2i\eta},
\end{eqnarray}
\begin{eqnarray}
Res(\beta) &=& \lim_{p'\to\beta}{d\over dp'}
\eta^{2} \left( p^{2} + p'^{2} + \eta^{2} \right) 
\left[ \frac{-(p'-\beta)}{(p'-\alpha)(p'-\alpha^{\ast})} +
\frac{1}{p'-\beta^{\ast}} \right]^{2} \\
&=& \eta^{2}\left\{ 2(p-i\eta)\left[ \frac{1}{2i\eta} \right]^{2} + 
4p(p-i\eta)\left[ \frac{1}{2i\eta} \right]
\left[ \frac{-1}{(2i\eta)^{2}} +
{1\over 2p(-2p+2i\eta)} \right]\right\} \\
&=& \frac{\eta^{2}}{2}\left\{ {p\over\eta^{2}} - {i\over\eta} +
{p^{2}\over i\eta^{3}} - {p\over\eta^{2}} - {1\over i\eta} \right\} \\
&=& {p^{2}\over 2i\eta}.
\end{eqnarray}
Hence the sum of residues is
\begin{eqnarray}
Res(\alpha)+Res(\beta) &=& {p^{2}\over 2i\eta} + {p^{2}\over 2i\eta}\\
&=& {p^{2}\over i\eta}
\end{eqnarray}
Since $(\eta^{2}/p'p)Q''_{0}(y)$ is symmetric around the origin, we
can integrate along the same contour as before and obtain
\begin{eqnarray}
\int^{\infty}_{0} \frac{\eta^{2}}{p'p} Q''_{0}(y)\, dp' &=&
{1\over 2} \oint \frac{\eta^{2}}{pz} Q''_{0}(y)\, dz \\
&=& {1\over 2} \left( 2\pi i \sum Res \right) \\
&=& \frac{\pi p^{2}}{\eta}.
\end{eqnarray}
From these results, it is obvious that Eq.~[\ref{eq:integral_2}] is
true.  The proof is complete.

A simple example is the momentum space Schr\"{o}dinger equation with a linear
potential in the $S$-state~\cite{maung93,norbury91},
\begin{equation}
\frac{p^{2}}{2\mu}\,\phi_{n0}(p)+\frac{\lambda_{L}}{\pi p^{2}}
\int^{\infty}_{0}\underbrace{\left[ {\eta^{2}\over p'p}Q''_{0}(y)+
Q'_{0}(y)\right]}_{V^{L}_{0}(p,p')}\,
\phi_{n0}(p')\, dp'
= E_{n0}\,\phi_{n0}(p),
\label{se1}
\end{equation}
where $y=(p^{2}+p'^{2})/2p'p$.  Lande 
subtraction~\cite{maung93,norbury91} involves subtracting a zero term
\begin{equation}
\int^{\infty}_{0}\left[ {\eta^{2}\over p'p}Q''_{0}(y)+Q'_{0}(y)\right]\,dp'
=0
\end{equation}
from Eq.~[\ref{se1}] such that
\begin{equation}
\frac{p^{2}}{2\mu}\,\phi_{n0}(p)+\frac{\lambda_{L}}{\pi p^{2}}\int^{\infty}_{0}
\left[ {\eta^{2}\over p'p}Q''_{0}(y)+Q'_{0}(y)\right]
[\phi_{n0}(p')-\phi_{n0}(p)]\, dp' 
= E_{n0}\,\phi_{n0}(p).
\label{se2}
\end{equation}
Using Eqs.~[\ref{eq:Q'},\ref{eq:Q''}], the integral in Eq.~[\ref{se2}] for 
$p>0$ in the limit of $y\to 1$ can be shown to equal
\begin{equation}
\lim_{\eta\to 0}\,\lim_{p\to p'}\,
{\lambda_{L}\over\pi}\,\left[2\eta^{2}\left({1\over (p-p')^{2}+\eta^{2}}
\right)^{2}
-{1\over (p-p')^{2}+\eta^{2}}\right]\,(p-p')^{2}\,{d\phi_{n0}\over dp}=0.
\label{sing1}
\end{equation}
The order of the limits in Eq.~[\ref{sing1}] is important.
The reverse order will lead to the nonsensical result $\int Q'_{0}(y)\,dp'=0$. 
Next, in the limit of $p,p'\to 0$, $(p+p')^{2}=(p-p)^{2}$.  
By substituting 
this equality into Eqs.~[\ref{eq:Q'},\ref{eq:Q''}], it can be shown again that
the integral in Eq.~[\ref{se2}] vanishes for $p\to 0$ at $y=1$.  At the end, 
the integral vanishes at $y=1,\;\forall\, p$.  Away from the singularities, 
both integrands in the integral of Eq.~[\ref{se2}] are finite.  By taking 
$\eta\to 0$, the first integrand vanishes.  The final form of Eq.~[\ref{se2}] 
is
\begin{equation}
\frac{p^{2}}{2\mu}\,\phi_{n0}(p)+\frac{\lambda_{L}}{\pi p^{2}}\int^{\infty}_{0}
Q'_{0}(y)\,[\phi_{n0}(p')-\phi_{n0}(p)]\, dp' 
= E_{n0}\,\phi_{n0}(p),
\label{lin0}
\end{equation}
where $Q'_{0}(y)=1/(1-y^{2})$.

The momentum space NRSE with a coulomb potential is given as
\begin{equation}
{p^{2}\over 2\mu}\phi_{nl}(p) + {\lambda_{C}\over \pi p}
\int^{\infty}_{0} P_{l}(y)\frac{Q_{0}(y)}{p'} \phi_{nl}(p')p'^{2}dp'
-{\lambda_{C}\over \pi p}\int^{\infty}_{0}w_{l-1}(y)\phi_{nl}(p')
p'dp'= E_{nl}\phi_{nl}(p).
\end{equation}
Use Eq.~[\ref{eq:integral_1}] to subtract out the logarithmic singularity
and obtain
\begin{eqnarray}
&& {p^{2}\over 2\mu}\phi_{nl}(p) + {\lambda_{C}\over \pi p}
\int^{\infty}_{0} P_{l}(y)\frac{Q_{0}(y)}{p'} \left[ p'^{2}\phi_{nl}(p')
- \frac{p^{2}\phi_{nl}(p)}{P_{l}(y)} \right] dp' +
{\lambda_{C}\over \pi p} \left[ {\pi^{2}\over 2}p^{2}\phi_{nl}(p)\right]
\nonumber \\
&& \quad -{\lambda_{C}\over \pi p}\int^{\infty}_{0}w_{l-1}(y)\phi_{nl}(p')
p'dp' = E_{nl}\phi_{nl}(p).
\end{eqnarray}

Before we perform Lande subtraction on the NRSE with a confining potential, we
need the identity
\begin{equation}
P'_{l}(1)=\frac{l(l+1)}{2}. \label{eq:P'}
\end{equation}
\noindent
Proof:\\
We use the recursion relation $xP'_{x}-P'_{l-1}(x)=lP_{l}(x)$, the
equality $P_{l}(1)=1$ to obtain the following relations:
\begin{eqnarray}
P'_{l}(1) - P'_{l-1}(1) &=& l \nonumber \\
P'_{l-1}(1) - P'_{l-2}(1) &=& l-1 \nonumber \\
\vdots\quad && \nonumber \\
P'_{2}(1) - P'_{1}(1) &=& 2 \nonumber \\
P'_{1}(1) - P'_{0}(1) &=& 1.
\end{eqnarray}
Add these relations and use $P_{0}(x)=1$ or $P'_{0}(x)=0$, we prove 
Eq.~[\ref{eq:P'}].
Next we want to examine the singularities of $Q'_{l}(y)$ and $Q''_{l}(y)$.
Differentiating Eq.~[\ref{eq:Ql}] once and twice, we have
\begin{eqnarray}
Q'_{l} &=& P'_{l}Q_{0} + P_{l}Q'_{0} - w'_{l-1}, \\
Q''_{l} &=& P''_{l}Q_{0} + 2P'_{l}Q'_{0} + P_{l}Q''_{0} - w''_{l-1}.
\end{eqnarray}
$\eta^{2}P''_{l}Q_{0}$, $\eta^{2}P'_{l}Q'_{0}$ and $\eta^{2}w''_{l-1}$
vanish in the limit of $\eta\to 0$.  The momentum space confining potential
is
\begin{equation}
V^{L}_{l}(p',p)={\lambda_{L}\over \pi} \lim_{\eta\to 0} \left\{ P_{l}(y)
\left[ \frac{\eta^{2}}{(p'p)^{3}} Q''_{0}(y) + \frac{Q'_{0}(y)}{(p'p)^{2}}
\right] + \frac{P'_{l}(y)Q_{0}(y) - w'_{l-1}(y)}{(p'p)^{2}} \right\}.
\end{equation}
Lastly, perform Lande subtraction and use Eq.~[\ref{eq:P'}].  Take the limit
of $\eta=0$, we derive the momentum space NRSE with a confining potential as
\begin{eqnarray}
&& {p^{2}\over 2\mu}\phi_{nl}(p) + {\lambda_{L}\over \pi p^{2}}
\int^{\infty}_{0} P_l(y)Q'_{0}(y) 
\left[ \phi_{nl}(p') - {\phi_{nl}(p) \over P_{l}(y)} \right]dp' \nonumber \\
&& \quad + {\lambda_{L}\over \pi p^{2}} \int^{\infty}_{0} P'_{l}(y)
{Q_{0}(y)\over p'} \left[ p'\phi_{nl}(p') - {l(l+1)\over 2}
\frac{p\phi_{nl}(p)}{P'_{l}(y)} \right]dp' \nonumber \\
&& \quad + {\lambda_{L}\over \pi p^{2}} {l(l+1)\over 2} \left[ {\pi^{2}\over 2}
p\phi_{nl}(p) \right]
+ {\lambda_{L}\over \pi p^{2}} \int^{\infty}_{0} w'_{l-1}(y)
\phi_{nl}(p')\,dp' = E_{nl}\phi_{nl}(p).
\end{eqnarray}

\section{Exact Solution of NRSE with a Linear Potential}

\begin{figure}[ht]
\begin{center}
\epsfig{file=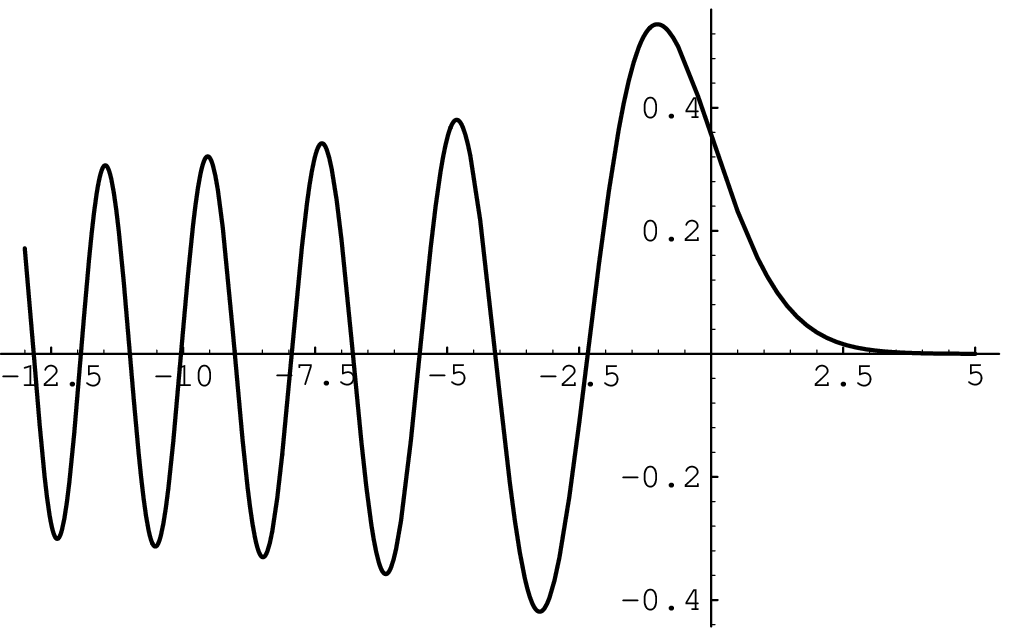,width=12cm,height=12cm}
\caption{\label{ai}
The graph of the Airy function--The zeros are all negative.  Since $S=r\,R$, 
$S$ must be
zero at the origin $r=0$.  The eigen-energy acts like a horizontal shift
which shifts the origin back to a zero.}
\end{center}
\end{figure}

In the next section, we are going to solve the NRSE numerically with a linear
potential.  In this section, we will solve the same equation exactly so that
we can use the analytic results to check their numerical counterparts.  The
Hamiltonian equation can be written as
\begin{equation}
\left( {d^{2}\over dr^{2}} + {2\over r}\,{d\over dr} \right)\,R -
2\mu[\lambda_{L}\,r - E]\,R = 0. \label{linear_r}
\end{equation}
Let $S\equiv r\,R$, then Eq.~[\ref{linear_r}] can be simplified as
\begin{equation}
{d^{2}\over dr^{2}}S - 2\mu[\lambda_{L}\,r - E]S = 0.
\end{equation}
If we define as new variable
\begin{equation}
x\equiv \left( {2\mu\over \lambda_{L}}^{2} \right)^{1\over 3}
[\lambda_{L}r - E], \label{linear_s}
\end{equation}
Eq.~[\ref{linear_s}] can be transformed as
\begin{equation}
S'' - xS = 0,
\end{equation}
which is the Airy equation.  The solution which satisfies the boundary
condition of $S\to 0$ as $x\to \infty$ is the Airy function
${\rm Ai}(x)$.  Fig.~[\ref{ai}] illustrates the graph of the
Airy function.  By noticing
that $S\equiv r\,R$ vanishes at $r=0$, we infer that $S(r=0)$ must 
coincide with a zero of ${\rm Ai}(x)$.  It is made possible by letting the 
eigen-energy act as a horizontal shift
which shifts the origin to the left along the $x$-axis.  If $S$ is plotted
against $r$ instead of $x$, $S$ will vanish at the origin if $E_{n}$
is chosen appropriately.  This conclusion leads to the 
eigen-energy formula
\begin{equation}
E_{n} = -x_{n}\,\left( {\lambda_{L}^{2}\over 2\mu} \right)^{1\over 3},
\end{equation}
where $x_{n}$ is the $n$-th zero of the Airy function counting from
$x=0$ along $-x$.

In Norbury \emph{et. al.}'s~\cite{norbury91} paper, the values
$\lambda_{L} = 5$ and $\mu = 0.75$ are used.  The eigen-energy formula is
\begin{equation}
E_{n} = -2.554364772\,x_{n}.
\end{equation}
Table~\ref{ai_tab} lists the zeros of the Airy function and the corresponding
exact eigen-energies.

\section{Conclusion}
The $p$-space formalism shown in this paper can be applied to any arbitrary 
potential in principle although only the cases of $n=-1,1$ in $r^{n}$ are
considered in this paper.  More complicated potentials require the calculation
of integrals involving other powers of $r$.  The potential involving
the coupling constant $\alpha(r)$
in QCD near the asymptotic freedom region is such an example.  In these cases,
we may need to reply on numerical integration to evaluate the integrals of the
Lande subtraction terms.

\newpage

\begin{table}[ht]
\caption{Zeros of the Airy function and the corresponding
eigen-energies in GeV for $l=0$, $\mu = 0.75 \rm\, GeV$, 
$\lambda_{L} = 5 \rm\, GeV$.}
\begin{center}
\vskip 10pt
\begin{tabular}{c r r }
\hline
n & $x_{n}$ & $E_{n}$ \\
\hline
1  &  -2.33810741 & 5.972379202 \\
2  & -4.08794944 & 10.44211404 \\
3  & -5.52055983 & 14.10152355  \\
4  & -6.78670809 & 17.33572806 \\
5  & -7.94413359 & 20.29221499 \\
6  & -9.02265085 & 23.04714148 \\
7  & -10.04017434 & 25.64626764 \\
8  & -11.00852430 & 28.11978666 \\
9  & -11.93601556 & 30.48893766 \\
10 & -12.82877675 & 32.76937540 \\
\hline
\end{tabular}
\end{center}
\label{ai_tab}
\end{table}

\newpage

\begin{thebibliography}{99}

\bibitem{iitaka}
T. Iitaka, \emph{Physical Review E}, {\bf 49}, 4684 (1994).

\bibitem{succi}
S. Succi, \emph{Physical Review E}, {\bf 53}, 1969 (1996).

\bibitem{maung93}
K. M. Maung, D. E. Kahana and J. W. Norbury, \emph{Physical
Review D}, {\bf 47}, 1182 (1993).

\bibitem{sak85}
J. J. Sakurai, {\em Modern Quantum Mechanics}, (Redwood City, CA: Addison-
Wesley, 1985), 238-242.

\bibitem{stegun}
M. Abramowitz and I. A. Stegun, {\em Handbook of Mathematical Functions},
(Washington, D. C.: National Bureau of Standards, 1964), 333--335.

\bibitem{GR80}
I. S. Gradshteyn and I. M. Ryshik, \emph{Tables of Integrals, Series, and
Products}, (San Diego: Academic, 1980), 526.

\bibitem{norbury91}
J. W. Norbury, D. E. Kahana and K. M. Maung, \emph{Canadian
Journal of Physics},  {\bf 70}, 86 (1991).

\bibitem{nr97}
W. H. Press, S. A. Teukolsky, W. T. Vetterling and B. P. Flannery,
{\emph Numerical Recipes in C}, (Cambridge: Cambridge, 1997).

\bibitem{gasi74}
S. Gasiorowicz, \emph{Quantum Physics} (New York: Wiley, 1974), 197.

\end{thebibliography}
\end{document}